# Family of Nonequilibrium Statistical Operators and the Influence of the Past on the Present

**V.V. Ryazanov**

Institute for Nuclear Research National Academy of Sciences of Ukraine,
47, Nauky Prosp., Kiev 03068, Ukraine

E-mail: vryazan@kinr.kiev.ua

A family of nonequilibrium statistical operators (NSO) is introduced which differ by the system lifetime distribution over which the quasiequilibrium distribution is averaged. This changes the form of the source in the Liouville equation, as well as the expressions for the kinetic coefficients, average fluxes, and kinetic equations obtained with use of NSO. The difference from the Zubarev form of NSO is of the order of the reciprocal lifetime of a system.

PACS numbers: 02.50.-r; 05.70.Ln; 05.40.+j; 31.70.Hq

*Keywords: Nonequilibrium Statistical Operator; Lifetime; Past.*

In work [1] the new interpretation of a method of the Nonequilibrium Statistical Operator (*NSO*) is given, in which instead of operation of a taking an invariant part [2] or of auxiliary "weight function " (in a terminology of works [4,5]) *NSO* is treated as averaging of the quasi-equilibrium statistical operator on the system past lifetime distribution. This approach is coordinated to operations developed for the general theory of random processes, in the renewal theory, with carried out of Zubarev in work [3] averaging on the initial moments of time by reception *NSO*, etc.

In works of Kirkwood it was marked, that the condition of system at the current moment of time depends on all previous evolution of nonequilibrium processes developing in it. In works [4,5] is specified, that it is possible much "of weight functions" to result. The arbitrary kind of lifetime density distribution enables to write down a general view of a source in the dynamic Liouville equation, which thus accepts Boltzmann-Prigogine form and contains dissipative effects. If in works [2,3] the linear form of a source received from exponential lifetime distribution is used, other expressions for the lifetime distributions give more complete and exact analogues "of integrals of collisions". The obvious account of choose of the direction of time (through finiteness of the lifetime, beginning, end and irreversibility of life) is entered. Except for connection with the theory of queues, the theory of reliability, theory of management, theory of the information etc., in offered work reflects many physical consequences connected to fundamental physical problems. The formalism follows from the physical contents, for example, from finiteness of the lifetime of real physical systems (can result many examples of tasks, in which it is necessary to consider systems of the finite sizes with the finite lifetime). Generally description of nonequilibrium systems represents the self-coordinated task: definition of the lifetime through interaction of system with an environment, dynamics of the operators describing nonequilibrium processes, and substitution of the found average lifetime in *NSO* and definition of the nonequilibrium physical characteristics dependent on the lifetime of the system.

In [1] the Nonequilibrium Statistical Operator introduced by Zubarev [2-3] rewritten as

$$ln\,\rho\,(t) = \int_0^\infty p_q(u) ln\,\rho_q(t-u,\,-u)\,du,$$

where $ln\,\rho\,(t)$ is the logarithm of the *NSO* in Zubarev's form, $ln\,\rho_q(t,\,0)$ is the logarithm of the quasi-equilibrium distribution. In [1] the auxiliary weight function $p_q(u) = \varepsilon\,exp\{-\varepsilon u\}$ was interpreted as the probability distribution density of the lifetime of a system from the moment $t_0$ of its birth till the current moment $t$. This time period can be called the time period of getting information about system from its past. Instead of the exponential distribution $p_q(u)$ any other sample distribution could be taken. This fact was marked in [1] and [4-5] (where the distribution density $p_q(u)$ is called auxiliary weight function). In [6] it was noted that the role of the form of the source term in the Liouville equation in *NSO* method has never been investigated. In [7] it is stated that the exponential distribution is the only one which possesses the Markovian property of the absence of contagion, that is whatever is the actual age of a system, the remaining time does not depend on the past and has the same distribution as the lifetime itself. It is known [2-5] that the Liouville equation for *NSO* contains the source $J = J_{zub} = -\varepsilon\,[ln\,\rho\,(t) - ln\,\rho_q(t,0)]$ which tends to zero after taking the thermodynamic limit and setting $\varepsilon \to 0$, which in the spirit of the paper [1] corresponds to the infinitely large lifetime value of an infinitely large system. If $\varepsilon^{-1} = \langle t-t_0 \rangle$ and $d\varepsilon/dt = -\varepsilon^2$ one should add to this source the term $-\varepsilon [\ln \rho(t) - \int \varepsilon^2 u \exp\{-\varepsilon u\} \ln \rho_q(t-u, -u) du]$ which appears if one takes into account the time dependence of $\varepsilon$. If one sets $\varepsilon = 2/\langle t_f - t_0 \rangle$ (where $t_f$ is the random time moment of finishing the lifetime of a system) then $d\varepsilon/dt = 0$, and we recover the Zubarev's value $J_{zub}$. For a system with finite size this source is not equal to zero. In [5] this term enters the modified Liouville operator and coincides with the form of Liouville equation suggested by I.Prigogine [8] (the Boltzmann-Prigogine symmetry), when the irreversibility is entered in the theory on the microscopic level. We note that the form of *NSO* by Zubarev cast in [1] corresponds to the main idea of [8] in which one sets to the distribution function $\rho$ ($\rho_q$ in Zubarev's approach) which evolves according to the classical mechanics laws, the coarse distribution function $\tilde{\rho}$ ($\rho(t)$ in the case of Zubarev's *NSO*) whose evolution is described probabilistically since one perform an averaging with the probability density $p_q(u)$. The same approach (but instead of the time averaging the spatial averaging was taken) was performed in [9].

Besides the Zubarev's form of *NSO* [2-3], Green-Mori form [10-11] is known, where one assumes the auxiliary weight function [5] to be equal $W(t,t`) = 1 - (t-t`)/\tau$; $w(t,t`) = dW(t,t`)/dt` = 1/\tau$; $\tau = t - t_0$. After averaging one sets $\tau \to \infty$. This situation at $p_q(u=t-t_0) = w(t, t`=t_0)$ coincides with the uniform lifetime distribution. The source in the Liouville equation takes the form $J = \ln \rho_q / \tau$. In [2] this form of *NSO* is compared to the Zubarev's form.

One could name many examples (at least 1000) of explicit defining of the function $p_q(u)$. Every definition implies some specific form of the source term $J$ in the Liouville equation, some specific form of the modified Liouville operator and *NSO*. Thus the family of *NSO* is defined. If the distribution $p_q(u)$ contains $n$ parameters, one could write $n$ equations to express them through the system parameters. On the other hand, they are expressed through the lifetime values. Thus the problem arises of the optimal choose of the $p_q(u)$ function and the form of *NSO*. In the present article we restrict ourselves to the one-parametric function $p_q(u)$.

In [7] it was shown that the exponential lifetime distribution $t_f - t_0$ ($t_f$, $t_0$ are random moments of system death and birth) at big $t$'s the "age" of a system $t-t_0$ tends to the exponential form. In Zubarev's *NSO* [2-3] the lifetime value $t-t_0$ to the current time $t$, which is a part of the total lifetime $t_f - t_0$, is considered, that is the influence of the past on the current moment is taken into account. The full lifetime distribution, as well as the "past" lifetime (i.e. time from the system birth $t_0$ till the current time $t$) need not be exponential. For example, one may choose the $p_q(u)$ function as

$$p_q(u) = \varepsilon^2 u \exp\{-\varepsilon u\}, \qquad u = t-t_0, \qquad (1)$$

that is in the form of gamma-distribution $p_q(u) = \varepsilon (\varepsilon u)^{k-1} \exp\{-\varepsilon u\}/\Gamma(k)$, ($\Gamma(k)$ is gamma-function) at $k=2$. In this case the distribution (1) coincides with the special Erlang distribution of the order *2* [12], if the "failure" (in our case - the moment $t$) occurs at the end of the second period [12], the past of the system contains of two independent etaps. The distribution function itself equals $P_q(x) = 1 - \exp\{-\varepsilon x\} - \varepsilon x \exp\{-\varepsilon x\}$, $p_q(u) = dP_q(u)/du$, in the contrast of the exponential distribution where $P_q(x) = 1 - \exp\{-\varepsilon x\}$. The behaviour of these two distributions differs essentially near zero (*Fig.1*). In the case (1) the system has low probability to die at small values of $u$, contrarily to the exponential distribution where this probability is maximal. Any system exist during at least some minimal time which is reflected in the distribution (1). The logarithm of *NSO* in the case (1) has the form

$$\ln \rho(t) = \int_0^\infty p_q(u) \ln \rho_q(t-u, -u) du = \int_0^\infty \varepsilon^2 u \exp\{-\varepsilon u\} \ln \rho_q(t-u, -u) du =$$

$$\ln \rho_q(t, 0) + \int_0^\infty \sigma(t-u, -u)(1+\varepsilon u) \exp\{-\varepsilon u\} du =$$

$$\ln \rho_{zub}(t) + \int_0^\infty \sigma(t-u, -u) \varepsilon u \exp\{-\varepsilon u\} du; \qquad \sigma(t-u, -u) = d\ln \rho_q(t-u, -u)/du, \qquad (2)$$

where $\ln \rho_{zub}(t) = \ln \rho_q(t, 0) + \int_0^\infty \sigma(t-u, -u) \exp\{-\varepsilon u\} du = \int_0^\infty \ln \rho_q(t-u, -u) \varepsilon \exp\{-\varepsilon u\} du$ is the Zubarev's form of the *NSO*, $\ln \rho_q(t-u, -u)$ is quasiequilibrium distribution; the first time argument indicates the time dependence of the values $F_m$, which are thermodynamically conjugated to the thermodynamic variables; the second time argument $t_2$ in $\rho_q(t_1, t_2)$ denotes the time dependence through the Heizenberg representation for dynamical variables $P_m$ from which $\rho_q(t, 0)$ can depend [1-5]. It is seen from (2) that the logarithm of the NSO has an additional term in comparison to the Zubarev's form. The source in the rhs of the Liouville equation (or dissipative part of the Liouville operator [5,8]) equal $J=-\varepsilon [ln \rho (t) – ln \rho_{zub}(t)]$, that is the system relaxes not towards $ln \rho_q(t, 0)$, like it is the case of Zubarev's NSO, but towards $ln \rho_{zub}(t)$. The value of $\varepsilon$ is expressed through $<\Gamma>=<t-t_0>=2\varepsilon^{-1}$ (if one takes into account the $\varepsilon$ dependence on $t$, then $J=-\varepsilon [ln \rho (t) – ln \rho_{zub}(t)]-\varepsilon [ln \rho (t) - \int_0^\infty \varepsilon^3 u^2 exp\{-\varepsilon u\} ln \rho_q(t-u,-u) du]$) and through the physical properties of the system from relations of the kind [1] $\int_0^\infty \varepsilon^2 u exp\{-\varepsilon u\} <S(t-u,-u)> du = <S(t,0)> = -<ln \rho_q(t,0)>$ (it can be obtained from (2) by averaging over $\rho_q$ since $<\sigma>_q=0$; $<S(t,0)>$ is the nonequilibrium entropy of the system [2-5]) or using the relations recording in [1].

From the expression (2) it is seen that introduced NSO contains amendments to the Zubarev's NSO [2-3]. The physical results obtained with use of (2) also contains additional terms in comparison to Zubarev's NSO. The additional terms describe the influence of the lifetime finiteness on the kinetic processes. The expressions for average fluxes [2] averaged over (2) have the form

$$<j^m(x)> = <j^m(x)>_{zub} + \sum_n \int \int_{-\infty}^t \varepsilon (t-t`) exp\{\varepsilon (t`-t)\} (j^m(x), j^n(x`, t`-t)) X_m(x`, t`) dt` dx`, \quad (3)$$

where $<j^m(x)>_{zub}=<j^m(x)>_l + \sum_n \int \int_{-\infty}^t exp\{\varepsilon (t`-t)\} (j^m(x), j^n(x`, t`-t) X_m(x`, t`) dt` dx`$ are fluxes in the form obtained by Zubarev [2], $j^n$ are flux operators, $X_m$ are corresponding thermodynamical forces; $(j^m(x), j^n(x`, t)) = \beta^{-1} \int_0^\beta <j^m(x) (j^n(x`, t, i\tau)-<j^n(x`, t)>_l)>_l d\tau$ are quantum time correlation functions, $j^n(x`, t, i\tau)=exp\{-\beta^{-1} A\tau\} j^n(x`, t) exp\{\beta^{-1} A\tau\}$. The collision integrals of the generalized kinetic equation [2] averaged over (2) have the amendments

$$S^{(2)}_{add}=-\hbar^{-2} \int_{-\infty}^0 dt \varepsilon t exp\{\varepsilon t\} <[H_l(t), [H_l, P_k]+i\hbar \sum_m P_m \partial S^{(1)}_k/\partial <P_m>]>_q \quad (4)$$

to Zubarev result [2]: 
$$S^{(2)} = -\hbar^{-2} \int_{-\infty}^{0} dt\, exp\{\varepsilon t\} \langle [H_1(t), [H_1, P_k] + i\hbar \sum_m P_m \partial S^{(1)}_k / \partial \langle P_m \rangle ] \rangle_q,$$
where the Hamiltonian of the system is $H = H_0 + H_1$, $H_1$ is the Hamiltonian of the interaction which contains the longtime correlations [5], $S^{(1)}_k = \langle [P_k, H_1] \rangle_q / i\hbar$. The same is valid for the generalized transport equations [2], kinetic coefficients etc. Thus the selfdiffusion coefficient (or, to be exact, its Laplace transform over time and space) obtained in [3] in the form

$$D(\omega, q) = q^{-2} \Phi(\omega, q) / [1 + \Phi(\omega, q)/(i\omega - \varepsilon)], \tag{5}$$

where $\Phi(\omega, q) = \int_0^\infty dt\, exp\{(i\omega - \varepsilon)t\} \int_0^\beta \langle \dot{n}_q \dot{n}_{-q}(-t + i\hbar\lambda) d\lambda \rangle / \int_0^\beta \langle n_q n_{-q}(i\hbar\lambda) \rangle d\lambda$, after use of (2) takes on the form

$$D(\omega, q) = q^{-2} [\Phi(\omega, q) + \varepsilon\, d\Phi(\omega, q)/d(i\omega)] / \{1 + \Phi(\omega, q)/(i\omega - \varepsilon) +$$
$$\varepsilon [d\Phi(\omega, q)/d(i\omega) - \Phi(\omega, q)/(i\omega - \varepsilon)]/(i\omega - \varepsilon)\}. \tag{6}$$

At $\varepsilon \to 0$, for infinitely large system in the thermodynamic limit this expression (6) coincides with (5) at $\varepsilon \to 0$ [3]. For finite size systems (as well as for the case $\omega \to 0$) the results differ. The use of the *NSO* in the form (2) for neutron kinetics allows more precise description of the neutrons scattering and of neutrons slowing-down.

If one chooses as $p_q(u)$ the special Erlang distribution with $k=2,3,4,...,n$ and $P_q(x) = 1 - exp\{-\varepsilon x\}(1 + \varepsilon x/1! + ... + (\varepsilon x)^{k-1}/(k-1)!)$; $\varepsilon = k/\langle \Gamma \rangle$ which depends only on the average value of $\langle \Gamma \rangle$, that is on the age of the system, in our case (if $k=1$ the result coincides with the exponential distribution) at $k=n$ we have:

$$ln\, \rho_{n,\varepsilon}(t) = ln\, \rho(t) = \int_0^\infty [\varepsilon (\varepsilon u)^{n-1}/(n-1)!] exp\{-\varepsilon u\} ln\, \rho_q(t-u, -u) du =$$

$$ln\, \rho_{n-1,\varepsilon}(t) + \int_0^\infty \sigma(t-u, -u)[(\varepsilon u)^{n-1}/(n-1)!] exp\{-\varepsilon u\} du =$$

$$ln\, \rho_{zub}(t) + \int_0^\infty \sigma(t-u, -u) [\varepsilon u/1! + ... + (\varepsilon u)^{n-1}/(n-1)!] exp\{-\varepsilon u\} du; \tag{7}$$

that is the amendment to $ln\, \rho_{zub}(t)$ contains $n-1$ terms, the iteration procedure is performed. As it is indicated in [12] in this case the failure occurs after $k$ stages, and the durations of those stages are independent random values distributed exponentially. Thus the multistage model of the system past is introduced. Nonequilibrium processes typically proceed on different stages, each characterized by a proper time scale [13]. In the distribution (1) the account for two stages is performed. Other distributions may account for some other peculiarities of the system past. For example, general Erlang distribution with the density of probability $\theta\,\rho_1 exp\{-\rho_1 x\}+(1-\theta)\rho_2 exp\{-\rho_2 x\}$ (by 2 stages) describe the situation when with the probability $\theta$ the failure (moment $t$) occurs on the first stage with the density of probability of nonfailire work $\rho_1 exp\{-\rho_1 x\}$, and with the probability $(1-\theta)$ on the second stage with the density of probability $\rho_2 exp\{-\rho_2 x\}$. These stages can be various phase states of matter, stationary and nonstationary states, for example, and so on. The corresponding amendments will enter the expressions for the fluxes, collision integral and kinetic coefficients. The expression for the source in the rhs of the Liouville equation for relation (7) has the form $J=-\varepsilon\,[ln\,\rho_{n,\varepsilon}(t) - ln\,\rho_{n-1,\varepsilon}(t)]$, that is $n$-distribution relaxes towards ($n$-$1$)-distribution (at $<\Gamma>=<t-t_0>$, $\varepsilon=n/<\Gamma>$ and if one takes into account the time dependence of $\varepsilon$ the source has the form $J=-\varepsilon\,[ln\,\rho_{n,\varepsilon}(t) - ln\,\rho_{n-1,\varepsilon}(t)] - \varepsilon\,[ln\,\rho_{n,\varepsilon}(t) - ln\,\rho_{n+1,\varepsilon}(t)]$).

Besides the special kinds of Erlang distributions with given integer values of $k=n$ (we remain thus within the family of one-parametric distributions) more general one, two-parametric gamma-distribution can be used, where the $k$ parameter can take on arbitrary values. In this case $<\Gamma>=k/\varepsilon$. Formally the situation is possible with $k<1$. Then the source terms tend to infinity since $(t-t_0)^{k-1}\,|_{t=t_0}\to\infty$ at $k<1$. This divergence can be overcome if we limit the value of $t-t_0$ setting its bottom threshold by the minimal lifetime $\Gamma_{min}$ that is if we replace the lower zero integration limit by $\Gamma_{min}$. Then the source expression acquires an additional term $((\varepsilon\,\Gamma_{min})^{k-1}/\Gamma(k))\varepsilon\,exp\{-\varepsilon\,\Gamma_{min}\}\,ln\rho_q(t-\Gamma_{min},-\Gamma_{min})$.

The generalizations of the results to more common (in the very general case arbitrary functions $F(x)=P_q(x)$) classes of distributions is performed in [7] with use of the renewal theory approaches. At big $t$ the distribution of $t-t_0$ converges towards $F_0(u)=\mu^{-1}\int_0^u [1-F(s)]ds$, where $F(s)$ is the distribution of the value $t_f-t_0$ with average $\mu$. In [7] the limit case for all distributions $F(x)$ is obtained, it is shown that the scaled random value $(t-t_0)/t$ at $t\to\infty$ converges to the limit distribution density $g_\alpha(x)=(sin\,\pi\alpha/\pi)x^\alpha(1-x)^{\alpha-1}$, $0<\alpha<1$, $x\in[0, 1]$, which is related to the full lifetime distribution functions $F(x)$ with regular "tails", that is $1-F(x)=x^{\alpha} L(x)$, $0<\alpha<1$, where $L(tx)/L(t) \to 1$ at $t \to \infty$. The average value $<\Gamma>/t=(\alpha-1)\sin\pi\alpha/\sin\pi(\alpha-1)$. Since $<\Gamma>/t=\delta$ is small value the values of $\alpha$ are close to unity and $\delta \approx \sin\alpha\pi/\pi$. If $\alpha \approx 1-\delta$, $\sin(1-\delta)\pi = \sin\pi\delta = \delta\pi - (\delta\pi)^3/3+... \approx \pi\delta$, we get an identity. At $\alpha \approx 1-\delta$ the distribution $g_{\alpha}(x) \approx \delta(1-x)^{-\delta}/x^{1-\delta}$ behaves similarly to $\varepsilon \exp\{-\varepsilon x\}$ at $\varepsilon \sim \delta$, differing essentially only at $x \to 0$. In this case the universal distribution is also characterized only by one parameter $\alpha$, but the limit situation $t \to \infty$ and tails of the distribution seem not to describe fully the influence of the past on the present, since the nearest time moments with conserved memory have bigger significance.

The use of some other explicit forms of the lifetime distributions as $p_q(u)$ (namely, logarithmic logistic distribution with $p_q(u)=k\rho^k u^{k-1}/[1+(u\rho)^k]^2$, complex exponential distribution, obtained if the parameter of the intensity of the exponential distribution itself is a random value [14] which gives the Pareto distribution $p_q(u)=k(k/\rho_0)^k/(u+k/\rho_0)^{k+1}$, and so on) makes us to state that the deviation of the distribution of $\ln\rho(t)$ obtained with use of those distributions from $\ln\rho_{zub}(t)$ is of the order $1/<\Gamma>$. That's why in the expressions (3-4), (6) the amendments to Zubarev's result are proportional to $\varepsilon \sim <\Gamma>$. This result is in accordance of the results of the theory of complex systems asymptotic coarse graining [15], according to which the lifetime distribution has the form $p_q(u)=\varepsilon \exp\{-\varepsilon u\}+\lambda\varphi_1(u)+\lambda^2\varphi_2(u)+...$, where the small parameter $\lambda$ in our case correspond to the value $1/<\Gamma>$. In the general case the parameter $\lambda$ is arbitrary.

Thus in big systems their state in the current time moment is influenced only by the existence of the past of the system, its duration, that is the age of a system, and the peculiarities of the system history have only minor influence. If the life was long, it does not matter for the current moment, what it was and how it was filled by events. For systems with big lifetimes we can leave only one term with $\exp\{-\varepsilon u\}$, that is to take into account only the age of the system and not its details, neglecting all other terms $\sim \varepsilon$. In the thermodynamic limit we set $\varepsilon \to 0$, $<\Gamma> \to \infty$ [2]. We note that there are systems (for example, reaction-diffusion systems, neutrons in multiplying surroundings etc) in which lifetimes are not related so closely to the sizes of systems but are essentially determined by external influences.

These results seem to be useful in, e.g., investigation of small lifetime systems where one should not neglect the value $<\Gamma>^{-1}$, or if one is interested in more detailed description of the initial periods of system evolution. This situation seems to be contrarily to that predicted by the theory of random evolutions [16] where the coarsing is performed. In our case the "decoarsing", that is more detailed but more cumbersome description is suggested. A number of results which follow from the interpretation of *NSO* and $p_q(u)$ as lifetime distribution density [1] can be obtained from the stochastic storage theory [17] and queuing theory. For example, in [17] the general result is stated that the random value describing the occupation period of a service system (that is lifetime in our terminology) has continuous distribution $p_q(u)=g(u, x)=x\,k(u-x, u)$, $u>x>0$; $g(u, x)=0$ otherwise, where $k(x, t)$ is the continuous distribution of the value $X(t)$ describing the input to the system.

Offered paper contains a number of new physical results, namely: 1). A physical basis of the recipes of reception of families *NSO* and their difference from Zubarev *NSO*. 2). Comparison of the approaches Zubarev and Prigogine. 3). The distribution of a form $\varepsilon^2 u\exp\{-\varepsilon u\}$ correctly describes behaviour of system at small times, existence of some minimal lifetime. 4). The additives to results of Zubarev in expressions for flows, generalized kinetic equations, generalized transfer equations, kinetic coefficients. 5). The special Erlang`s distribution describes real-life and important in the nonequilibrium description (as shown Bogoliubov) some stages in evolution of system. 6). The general Erlang`s distribution describes occurring in system and having a place in its past bifurcations and phase transitions. 7). The additives to Zubarev results are of the order of the reciprocal lifetime of a system and are essential at early stages of evolution of system and to systems with small lifetime. Chosen of Zubarev the form of distribution for the lifetime represents limiting distribution [15]. 8). Limiting universal density of distribution for the lifetime with power dependence is considered, and is shown, that she not full describes influence of the past on the present, as it is fair only for the large times. 9). The choice of distribution of the lifetime in *NSO* is connected in view of influence of the past of system, his physical features, on the present moment, for example, to the account only of age of system, as in Zubarev NSO, or with more detailed characteristic of past evolution of the system.

# References

[1] V.V. Ryazanov, Fortschritte der Phusik/Progress of Physics, **49** (2001) 885.

[2] D.N. Zubarev, *Nonequilibrium statistical thermodynamics*, Plenum-Consultants Bureau, 1974.

[3] D.N. Zubarev, in *Reviews of Science and Technology: Modern Problems of Mathematics*. Vol.15, pp. 131-226, (in Russian) ed. by R. B. Gamkrelidze, (Izd. Nauka, Moscow, 1980) [English Transl.: J. Soviet Math. **16**, 1509 (1981)].

[4] A.R.Vasconcellos, R.Luzzi, L.S.Garcia-Colin, Phys. Rev. A **43** (1991) 6622; Phys. Rev. A **43** (1991) 6663.

[5] J.G. Ramos, A.R. Vasconcellos and R. Luzzi, Fortschr. Phys./Progr. Phys. **43** (1995)265.

[6] V.G. Morozov, G.Röpke, Condensed Matter Physics, **1** (1998) 673.

[7] W. Feller, *An Introduction to Probability Theory and its Applications*, vol.2, J.Wiley, 1971.

[8] I. Prigogine, *From Being to Becoming* , Freeman, 1980.

[9] Yu.L. Klimontovich, *Statistical Theory of Open Systems*, Kluwer Acad. Publ., 1995.

[10] M.S. Green, J. Chem. Phys. 20 (1952)1281; ibid. 22 (1954) 398.

[11] H. Mori, I. Oppenheim and J. Ross, in *Studies in Statistical Mechanics* I, edited by J.de Boer and G.E. Uhlenbeck, North-Holland, 1962, pp. 217-298.

[12] D.R. Cox, *Renewal theory* , Methuen; John Wiley, 1961.

[13] N.N. Bogoliubov, in *Studies in Statistical Mechanics* I, edited by J. de Boer and G.E. Uhlenbeck, North Holland, 1962, pp. 4-118.

[14] D.R. Cox and D. Oakes, *Analysis of Survival Data* ,Chapman and Hall, 1984.

[15] V.S. Korolyuk and A.F. Turbin, *Mathematical Foundations of the State Lumping of Large Systems*, Kluwer Acad.Publ., 1993.

[16] V.S. Korolyuk, A.V. Swishuk, *Semi-Markov Random Evolutions*, Naukova dumka, 1992 (InRussian).

[17] N.U. Prabhu, *Stochastic Storage Processes*, Springer, 1980.

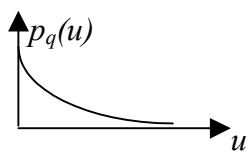 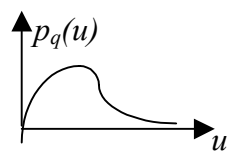

*Fig.1.a*  *Fig.1.б*

*Fig.1.a: exponential distribution, κ=1.*
*Fig.1.б: Gamma-distribution, with κ=2.*